\documentclass[aps,prl,twocolumn,superscriptaddress,longbibliography,floatfix]{revtex4-1}
\usepackage[colorlinks=true,allcolors=blue, bookmarks=false]{hyperref}
\usepackage{amsfonts,amsmath,amssymb}
\usepackage{graphicx}

\DeclareMathOperator{\Tr}{Tr}

\begin{document}

\title{Random Matrix Ensemble for the Level Statistics of Many-Body Localization}

\author{Wouter Buijsman} 

\affiliation{Institute for Theoretical Physics Amsterdam and Delta Institute for Theoretical Physics, University of Amsterdam, Science Park 904, 1098 XH Amsterdam, The Netherlands}

\email{w.buijsman@uva.nl}

\author{Vadim Cheianov}

\affiliation{Instituut-Lorentz and Delta Institute for Theoretical Physics, Universiteit Leiden, P.O. Box 9506, 2300 RA Leiden, The Netherlands}

\author{Vladimir Gritsev}

\affiliation{Institute for Theoretical Physics Amsterdam and Delta Institute for Theoretical Physics, University of Amsterdam, Science Park 904, 1098 XH Amsterdam, The Netherlands}

\affiliation{Russian Quantum Center, Skolkovo, Moscow 143025, Russia}

\date{\today}

\begin{abstract}
We numerically study the level statistics of the Gaussian $\beta$ ensemble. These statistics generalize Wigner-Dyson level statistics from the discrete set of Dyson indices $\beta = 1,2,4$ to the continuous range $0 < \beta < \infty$. The Gaussian $\beta$ ensemble covers Poissonian level statistics for $\beta \to 0$, and provides a smooth interpolation between Poissonian and Wigner-Dyson level statistics. We establish the physical relevance of the level statistics of the Gaussian $\beta$ ensemble by showing near-perfect agreement with the level statistics of a paradigmatic model in studies on many-body localization over the entire crossover range from the thermal to the many-body localized phase. In addition, we show similar agreement for a related Hamiltonian with broken time-reversal symmetry.
\end{abstract}

\maketitle

Random matrix theory \cite{Mehta04, Guhr98} provides an essential toolbox in nuclear \cite{Mitchell88, Weidemuller09, Dietz17}, condensed matter \cite{Shklovskii93, Oganesyan07, Serbyn16} and mesoscopic \cite{Beenakker93, Beenakker13} physics, and is used as well in \emph{e.g.} high energy physics \cite{Giordano14, Li17, Kovacs18}. In these fields, the physical interest for random matrix theory comes from the apparent universality of the local spectral statistics of quantum systems that are chaotic in the semiclassical limit \cite{Bohigas84}. Inspired by seminal works of Wigner \cite{Wigner55} and Dyson \cite{Dyson62}, one typically compares local spectral statistics with the local eigenvalue statistics of random matrices taken from the Gaussian orthogonal (GOE), unitary (GUE), or symplectic (GSE) ensemble -- depending on the type of transformation by which the Hamiltonian is diagonalized. These so-called Wigner-Dyson level statistics provide an excellent description of the local spectral statistics of a vast majority of the systems that are considered as quantum chaotic (ergodic) \cite{Poilblanc93}.

The GOE, GUE, and GSE are covered by to the Gaussian $\beta$ ensemble \cite{Dyson62, Forrester10}. Here, $\beta \in (0, \infty)$ is a continuous parameter which for the GOE, GUE, and GSE corresponds to $\beta = 1,2,4$, respectively. The Gaussian $\beta$ ensemble also covers Poissonian level statistics ($\beta \to 0$), as typically observed for regular (non-ergodic) systems \cite{Berry77, Relano04}. The Gaussian $\beta$ ensemble provides a smooth interpolation between Poissonian and Wigner-Dyson level statistics. Thanks to relatively recent progress made by Dumitriu and Edelman \cite{Dumitriu02}, the eigenvalue statistics of the Gaussian $\beta$ ensemble can be sampled at low computational costs. 

Physical systems displaying level statistics that can be tuned from Poissonian to Wigner-Dyson are of central interest in the field of many-body localization (MBL) \cite{Basko06, Abanin17}. Numerical studies provide evidence \cite{BarLev15, Agarwal15, Khemani17, Luitz17} for an intermediate phase characterized by \emph{e.g.} Griffiths effects in between the thermal (corresponding to $\beta \approx 1$) and the MBL (corresponding to $\beta \approx 0$) phase at finite system sizes. In this work, we numerically study the level statistics of a standard model in studies on MBL. Remarkably, we find near-perfect agreement with the eigenvalue statistics of the Gaussian $\beta$ ensemble over the entire crossover range, where $\beta$ is a single fitting parameter. Additionally, we show that similar agreement holds for a related Hamiltonian with broken time-reversal symmetry. We interpret the eigenvalue statistics of the Gaussian $\beta$ ensemble as generalized Wigner-Dyson level statistics. We show how the Gaussian $\beta$ ensemble provides a smooth interpolation between Poissonian and Wigner-Dyson level statistics by a systematic investigation of the eigenvalue statistics for $\beta \in [0,1]$.

\textit{Gaussian $\beta$ ensemble}.--- 
An ensemble of random matrices $T$ is described by a probability distribution $P(T)$ \cite{Mehta04}. An example is the GOE. For this ensemble of real symmetric matrices, the probability distribution is given by
\begin{equation}
P(T) = C_n e^{- \Tr \left( T^2 \right)}
\end{equation}
where $C_n$ is a normalization constant and $\Tr(\cdot)$ denotes a trace. The GOE is invariant under transformations $T \to O^{-1} T O$ for real orthogonal matrices $O$. Similarly, the GUE and GSE are invariant under unitary and symplectic transformations, respectively. Because there are only three types of associative division algebras (real, complex, and quaternionic numbers), no invariant random matrix ensembles exist beyond the GOE, GUE, and GSE. The joint probability distribution for the eigenvalues $\{ e_i \}$ of $n$-dimensional matrices from the Gaussian ensembles is given by
\begin{equation}
\rho(e_1, \ldots, e_n) = C_{\beta,n} \prod_{i < j} |e_i - e_j |^\beta \prod_{i=1}^n e^{- \frac{\beta}{2} e_i^2},
\label{eq: rho}
\end{equation}
where $C_{\beta, n}$ is a known normalization constant. As mentioned above, the Dyson index $\beta$ is given by $\beta = 1,2,4$ for the GOE, GUE, and GSE, respectively. 

An interpolation between the eigenvalue statistics of the invariant ensembles is provided by the Gaussian $\beta$ ensemble \cite{Dyson62, Forrester10}. This ensemble has a joint eigenvalue distribution given by Eq. \eqref{eq: rho} for the continuous parameter $\beta \in (0, \infty)$.  It was found only relatively recently \cite{Dumitriu02} that the eigenvalues of the tridiagonal matrix ensemble
\begin{equation}
T = \frac{1}{\sqrt{\beta}}
\begin{bmatrix}
a_n		& b_{n-1} \\
b_{n-1}	& a_{n-1}	& b_{n-2} \\
		& b_{n-2}	& a_{n-2}	& b_{n-3} \\
		& \ddots		& \ddots		& \ddots \\
		&			& b_2		& a_2		& b_1 \\
		&			&			& b_1		& a_1
\end{bmatrix}
\label{eq: T}
\end{equation}
with $a_i$ distributed according to the standard Gaussian distribution, for which the probability density is given by
\begin{equation}
P(a_i) = \frac{1}{\sqrt{2 \pi}} e^{- a_i^2 / 2},
\end{equation}
and $b_i$ distributed according to the $\chi$ distribution with the shape parameter given by $i \beta$, for which the probability density is given by
 \begin{equation}
 P(b_i) = 
 \begin{cases}
0 & \text{if } b_i \le 0, \\
\frac{2}{\Gamma(i \beta / 2)} b_i^{i \beta - 1} e^{- b_i^2} & \text{if } b_i > 0,
\end{cases}
 \end{equation} 
are distributed according to Eq. \eqref{eq: rho}. This matrix ensemble has the property that the eigendistribution factorizes into separate terms for the eigenvalues and the eigenvectors. Eq. \eqref{eq: T} allows one to sample from the Gaussian $\beta$ ensemble at low computational costs, and thus to generalize Wigner-Dyson level statistics beyond $\beta = 1,2,4$. Various aspects of the Gaussian $\beta$ ensemble have been studied in mathematical \cite{Dumitriu06, Forrester10} and physical \cite{LeCaer07, Relano08, Vivo08} contexts.

First, we study the eigenvalue statistics of the Gaussian $\beta$ ensemble for $\beta \in [0,1]$ by focusing on two common statistical measures: the distribution of the ratios of consecutive level spacings \cite{Oganesyan07, Atas13} and the level spacing distribution \cite{Mehta04}. For a set of eigenvalues $\{ e_i \}$ sorted in ascending order, the level spacings $\{ s_i \}$ are given by $s_i = e_{i+1} - e_i$, and the ratios $\{ r_i \}$ of consecutive level spacings are given by
\begin{equation}
r_i = \min \left( \frac{s_{i+1}}{s_i}, \frac{s_i}{s_{i+1}} \right).
\label{eq: r}
\end{equation}
For Poissonian level statistics ($\beta = 0$), the level spacing distribution is given by $P(s) = \exp(-s)$, where the spacings have been rescaled such that $\langle s \rangle = 1$. Correspondingly, the distribution of $r \in [0,1]$ is given by $P(r) = 2/(1+r)^2$, with $\langle r \rangle = 2 \ln(2) - 1 \approx 0.386$. For $\beta > 0$, we obtain data by numerically diagonalizing matrices $T$ as given in Eq. \eqref{eq: T} of dimension $n = 10^5$. We determine the $100$ eigenvalues closest to zero for each realization, accumulating at least $10^6$ eigenvalues. Aiming to maximize the accuracy of the results, we unfold \cite{Haake10} data before analysis. For $n \to \infty$, the density of states is given by a semicircle with radius $2 \sqrt{n}$. This asymptotic result, which we use here to unfold data sampled from the Gaussian $\beta$ ensemble, serves as a good approximation at finite ($n \gtrsim 100$) values of $n$ \cite{Dumitriu06}.

\begin{figure}
\includegraphics[width=0.7\columnwidth]{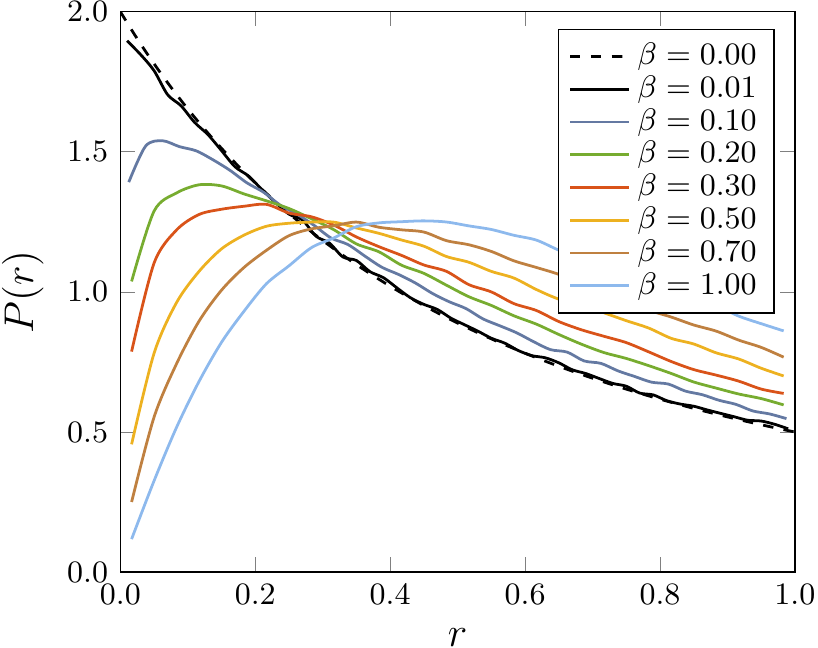} 

\smallskip

\includegraphics[width=0.7\columnwidth]{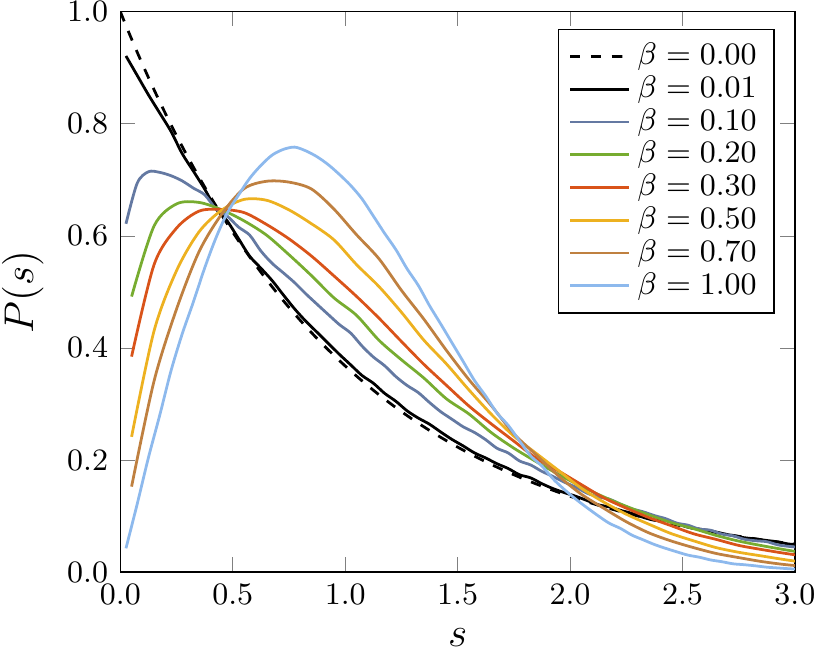} 
\caption{Numerically obtained distributons of $r$ (top) and $s$ (bottom) for the Gaussian $\beta$ ensemble at various $\beta$. The curves for $\beta = 0$ are obtained analytically from the expressions given in the main text.}
\label{fig: evaluation}
\end{figure}

Fig. \ref{fig: evaluation} shows the distributions of $r$ and $s$ for the Gaussian $\beta$ ensemble at various values of $\beta \in [0,1]$, indicating how the Gaussian $\beta$ ensemble interpolates between Poissonian and Wigner-Dyson level statistics. Table \ref{tab: rav} shows the average $\langle r \rangle$ as a function of $\beta$, which will be used as the fitting parameter when comparing the eigenvalue statistics of the Gaussian $\beta$ ensemble with the level statistics of a physical system.

\begin{table}
\centering
\begin{tabular}{c || c | c | c | c | c | c}
\hline \hline
$\beta$				& $0.00$		& $0.01$		& $0.05$		& $0.10$		& $0.15$		& $0.20$ \\ \hline
$\langle r \rangle$ 	& $0.386(3)$ 	& $0.389(0)$ 	& $0.398(4)$ 	& $0.408(9)$ 	& $0.420(1)$ 	& $0.429(5)$
\end{tabular}

\smallskip

\begin{tabular}{c || c | c | c | c | c | c}
$\beta$				& $0.25$		& $0.30$		& $0.35$		& $0.40$		& $0.45$		& $0.50$ \\ \hline
$\langle r \rangle$ 	& $0.438(1)$ 	& $0.446(2)$ 	& $0.453(6)$ 	& $0.461(7)$ 	& $0.469(3)$ 	& $0.475(8)$
\end{tabular}

\smallskip

\begin{tabular}{c || c | c | c | c | c | c}
$\beta$				& $0.55$		& $0.60$		& $0.65$		& $0.70$		& $0.75$		& $0.80$ \\ \hline
$\langle r \rangle$ 	& $0.482(6)$ 	& $0.489(0)$ 	& $0.494(6)$ 	& $0.500(8)$ 	& $0.505(8)$ 	& $0.511(2)$
\end{tabular}

\smallskip

\begin{tabular}{c || c | c | c | c | c | c}
$\beta$				& $0.85$		& $0.90$		& $0.95$		& $1.00$		& $2.00$		& $4.00$ \\ \hline
$\langle r \rangle$ 	& $0.516(4)$ 	& $0.521(5)$ 	& $0.526(2)$ 	& $0.530(2)$ 	& $0.599(7)$ 	& $0.673(9)$ \\
\hline \hline
\end{tabular}
\caption{Numerically obtained values of $\langle r \rangle$ for the Gaussian $\beta$ ensemble at various $\beta$ (see main text for details). The value for $\beta = 0$ is obtained from the expression given in the main text.}
\label{tab: rav}
\end{table}

\textit{Comparison with spectral statistics}.--- 
Here, we compare the level statistics of a standard model in studies on MBL with the eigenvalue statistics of the Gaussian $\beta$ ensemble. We consider a disordered spin-$1/2$ XXZ chain, for which the Hamiltonian is given by
\begin{equation}
H = \sum_{i=1}^{L} \left( S_i^x S_{i+1}^x + S_i^y S_{i+1}^y + \Delta S_i^z S_{i+1}^z \right) + \sum_{i=1}^L h_i S_i^z
\label{eq: H}
\end{equation}
with $S^\alpha_i =\frac{1}{2} \sigma^\alpha_i$, where $\sigma_i^\alpha$ are Pauli matrices ($\alpha =x,y,z$) acting on site $i$. During the last decade, the level statistics of this Hamiltonian have been studied extensively in \emph{e.g.} Refs. \cite{Oganesyan07, Pal10, Luitz15, Serbyn16, Bertrand16, Kjall18, Kudo18}. In particular, the intermediate level statistics between the thermal and the MBL phase have been studied by means of a two-stage flow picture in Ref. \cite{Serbyn16}. Following these references, we impose periodic boundary conditions $\sigma_{i+L}^\alpha \equiv \sigma_i^\alpha$, sample $h_i$ from the uniform distribution ranging over $[-W,W]$, set $\Delta = 1$ (unless stated otherwise), and restrict the focus to the symmetry sector $\sum_i S_i^z = 0$. We set $L=16$, for which $\dim(H)=12,870$. We consider at least $1000$ disorder realizations for each value of $W$. For each value of $W$ separately, we restrict the focus to the energy window containing the middle $10 \%$ of the union of all sampled spectra. The system exhibits a smooth crossover from Poissonian to Wigner-Dyson level statistics in the region $1.7 \lesssim W \lesssim 4.0$.

\begin{figure}
\includegraphics[width=0.7\columnwidth]{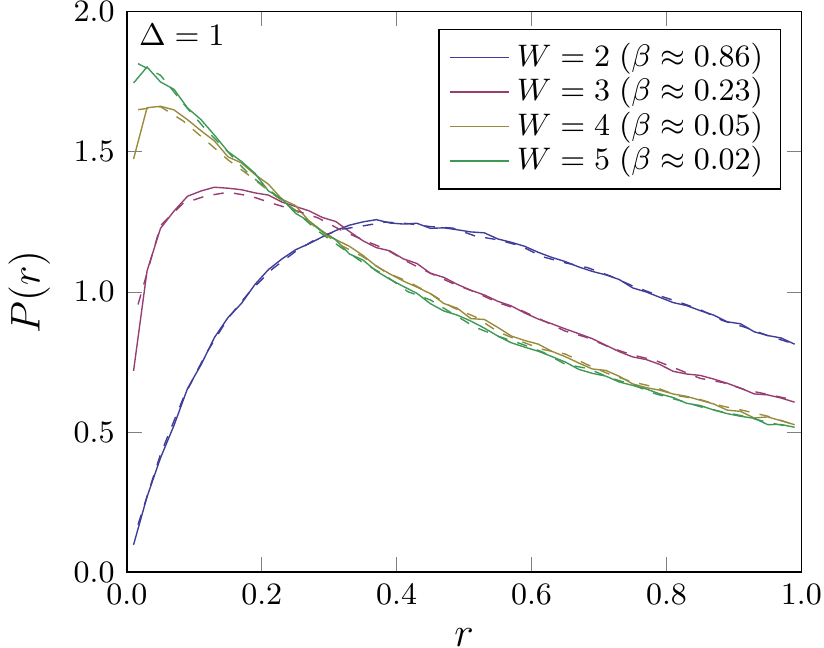} 

\smallskip

\includegraphics[width=0.49\columnwidth]{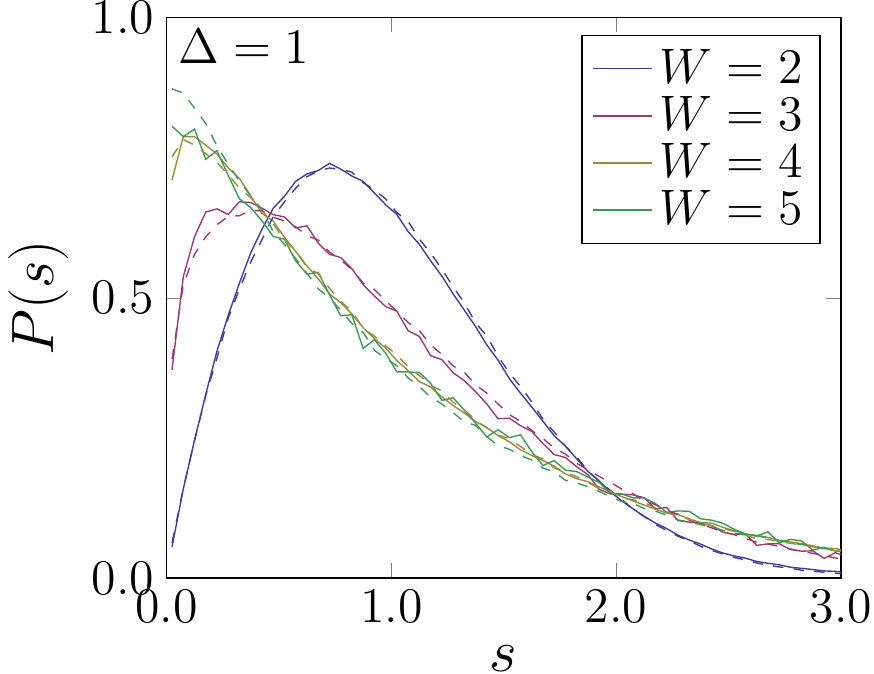} 
\includegraphics[width=0.49\columnwidth]{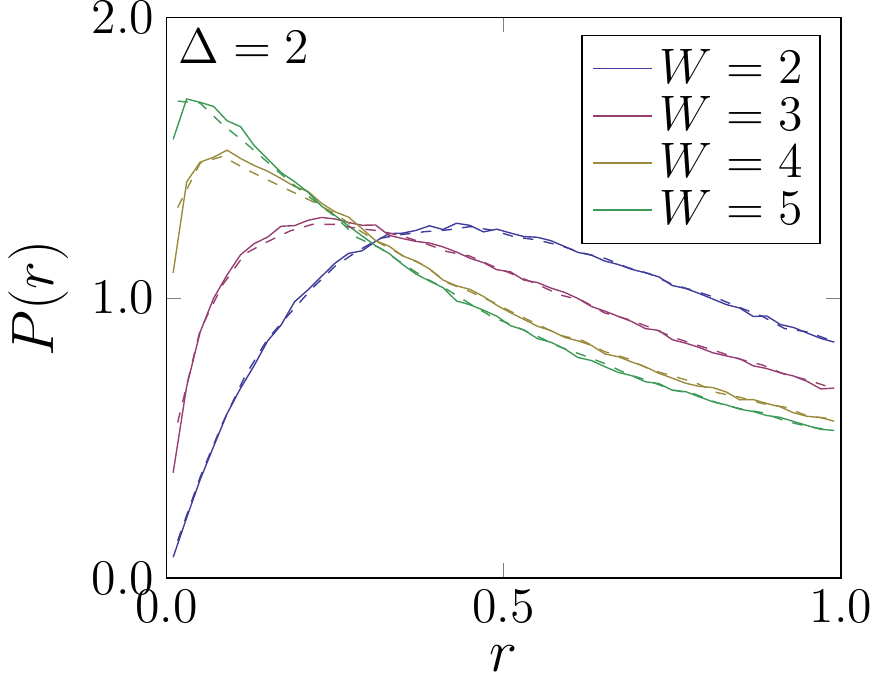} 
\caption{Numerically obtained distributions of $r$ and $s$ for the Hamiltonian at various $W$ (solid lines) and the corresponding distributions for the Gaussian $\beta$ ensemble (dashed lines, identical color scheme). The top and bottom left plots are for $\Delta = 1$, the bottom right one for $\Delta = 2$.}
\label{fig: validation}
\end{figure}

Fig. \ref{fig: validation} shows the distributions of $r$ and $s$ for the spectra of the Hamiltonian compared with the corresponding distributions for the Gaussian $\beta$ ensemble, where $\beta$ is estimated from $\langle r \rangle$. Note that, since $r$ is independent of the average level spacing, no unfolding \cite{Haake10} is required for drawing the distribution of this quantity. Before drawing the histograms of $s$ for the Hamiltonian, the spectra are unfolded by numerically estimating the smooth part of the density of states \footnote{We estimate the (scaled) smooth part of the density of states for each spectrum separately by using the Wolfram Research, Inc. \emph{Mathematica} version 10.0 function \texttt{SmoothKernelDistribution} with bandwidth specifications \texttt{\{"Adaptive", Automatic, 1\}, "Biweight"} ($W=2,3$) or \texttt{\{"Adaptive", Automatic, 1\}} ($W=4,5$).}. Remarkably, we observe near-perfect agreement between the spectral statistics of the Hamiltonian and the corresponding eigenvalue statistics of the Gaussian $\beta$ ensemble at all disorder strengths. Similar agreement can be found for $\Delta = 2$, which is illustrated in the lower right panel.

\begin{figure}
\includegraphics[width=0.49\columnwidth]{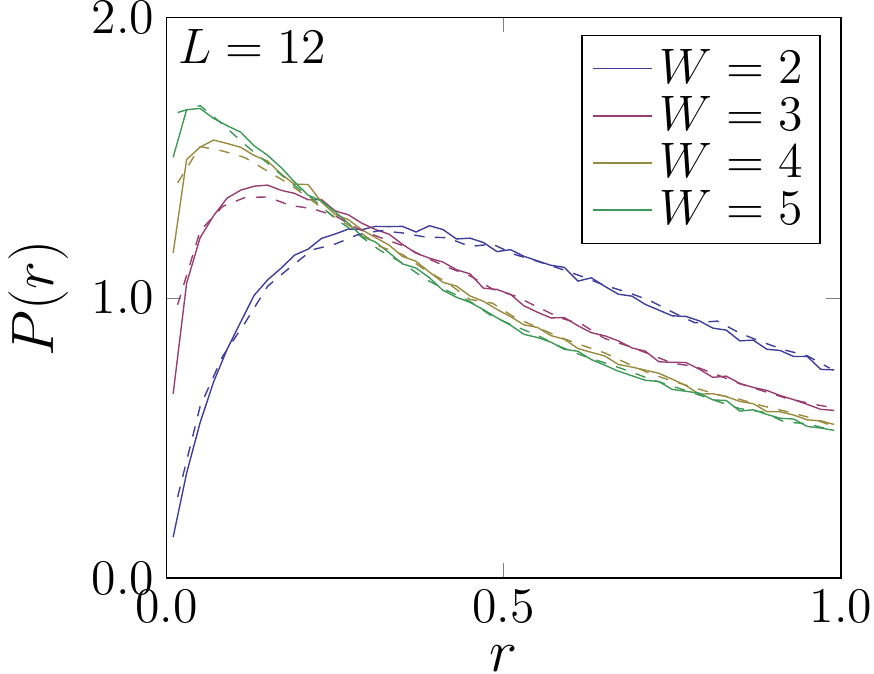}
\includegraphics[width=0.49\columnwidth]{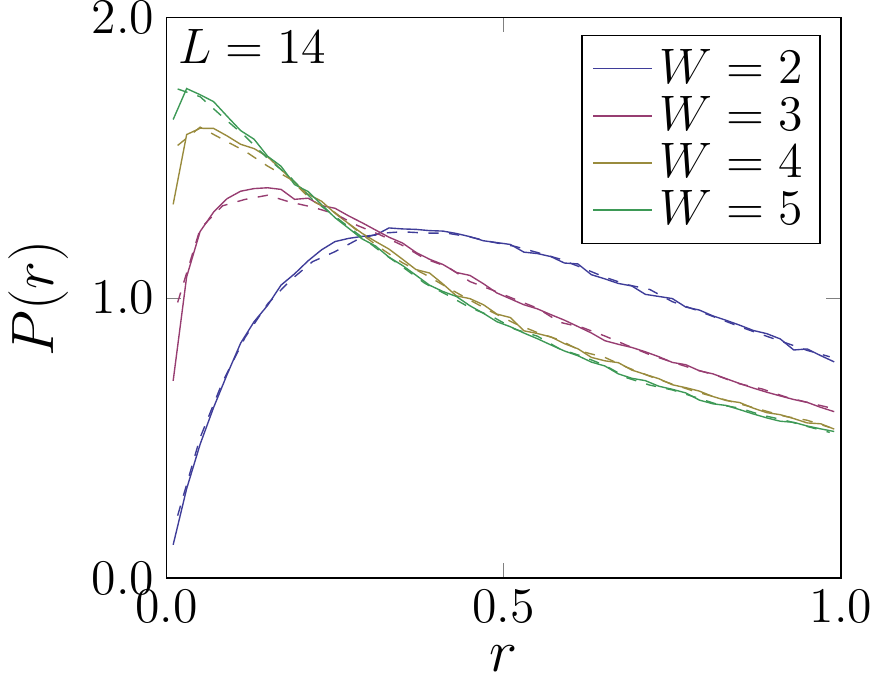} 

\smallskip

\includegraphics[width=0.49\columnwidth]{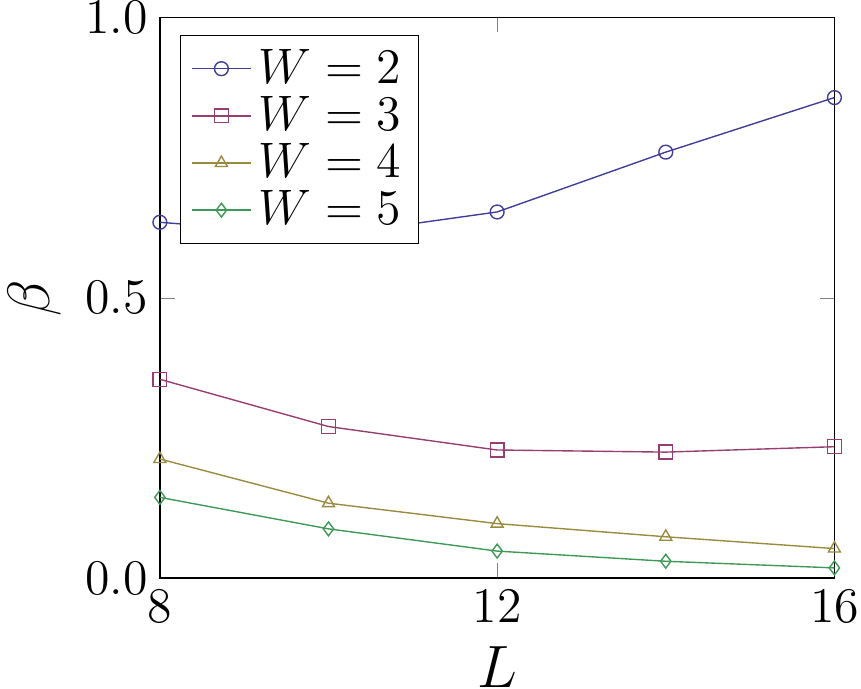}
\includegraphics[width=0.49\columnwidth]{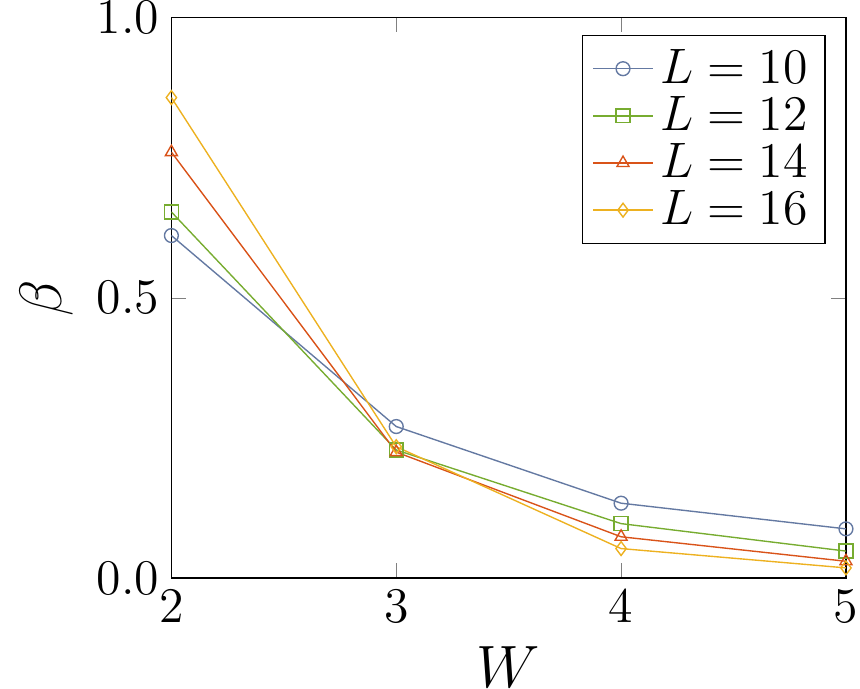} 
\caption{Numerically obtained distributions of $r$ for the Hamiltonian at various $W$ (solid lines) and the corresponding distributions for the Gaussian $\beta$ ensemble (dashed lines, identical color scheme) for $L=12,14$ (top panels) and the estimated value of $\beta$ for the spectra of the Hamiltonian as a function of $L$ and $W$ (bottom panels).}
\label{fig: finitesize}
\end{figure}

In Fig. \ref{fig: finitesize}, we study the sensitivity to finite-size effects. The top panels show that the agreement between the level statistics of the Hamiltonian and the Gaussian $\beta$ is near perfect also at $L=12,14$. The bottom panels show a flow towards Wigner-Dyson (Poissonian) level statistics for $W \lesssim 3$ ($W \gtrsim 3$) with increasing system size. As there is a one-to-one relation between $\langle r \rangle$ and $\beta$, these results can in principle be appended with previous results from \emph{e.g.} Ref. \cite{Luitz15}. Studying $\langle r \rangle$ as a function of the matrix dimension $n$ for the Gaussian $\beta$ ensemble indicates a difference of less than $1 \%$ between the value for $n=500$ and $n=10^5$ at all values $\beta \in [0,1]$.

\begin{figure}
\includegraphics[width=0.7\columnwidth]{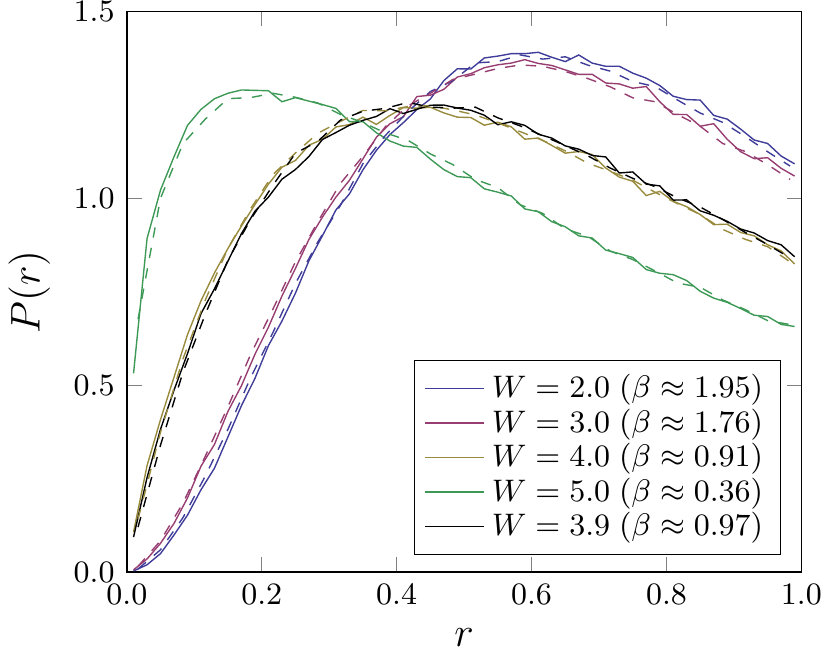} 
\caption{Numerically obtained distribution of $r$ for $H + H'$ at various $W$ (solid lines) and the corresponding distributions for the Gaussian $\beta$ ensemble (dashed lines, identical color scheme).}
\label{fig: brokensym}
\end{figure}

\newpage
\textit{Breaking time-reversal symmetry}.--- 
Ergodic systems with broken time-reversal symmetry are characterized by Wigner-Dyson level statistics for $\beta =2$ \cite{Mehta04}. For the Hamiltonian given in eq. \eqref{eq: H}, time-reversal symmetry can be broken in an experimentally relevant way by adding the $3$-body term
\begin{equation}
H' = \sum_{i=1}^L \vec{S}_i \cdot [\vec{S}_{i+1} \times \vec{S}_{i+2}],
\end{equation}
where $\vec{S_i} = [S_i^x, S_i^y, S_i^z]^T$ \cite{Avishai02}. For $\beta \gtrsim 1$, the distribution of the ratio of consecutive level spacings for the Gaussian $\beta$ ensemble can be approximated with high precision \cite{Atas13} from Eq. \eqref{eq: rho} with $n=3$, giving
\begin{equation}
P(r) \sim \frac{(r + r^2)^ \beta}{(1 + r + r^ 2)^{1 + 3 \beta /2}}.
\label{eq: Atas}
\end{equation}
In what follows, estimates of $\beta \ge 1$ from $\langle r \rangle$ are obtained by using Eq. \eqref{eq: Atas}. Fig. \ref{fig: brokensym} shows the distribution of $r$ for the spectra of $H + H'$ at several values of $W$ compared with the corresponding distributions for the Gaussian $\beta$ ensemble, where $\beta$ is estimated from $\langle r \rangle$. Again, we observe near-perfect agreement between correspondong curves at all disorder strengths.

\textit{Higher order spacing ratios}.--- 
Going beyond the study of the distribution of the ratios of consecutive level spacings and the level spacing distribution, we here study the higher order ratios $r^{(n)} \in [0,1]$ of level spacings, for a spectrum $\{ E_i \}$ sorted in ascending order defined as
\begin{equation}
r^{(n)}_i = \min \left( \frac{E_{i+2n} - E_{i+n}}{E_{i+n} - E_i}, \frac{E_{i+n} - E_i}{E_{i+2n} - E_{i+n}} \right).
\end{equation}
Note that $r^{(1)} = r$. For the Gaussian $\beta$ ensemble, it can be shown rigorously that the distribution of $r^{(n)}$ for $\beta = 2/(n+1)$ is equivalent to the distribution of $r^{(1)}$ for $\beta = 2(n+1)$ \cite{Forrester09}. Evidence for a broader class of inter-relations involving $\beta = 1,2,4$ has been provided recently in Ref. \cite{Tekur18}.

\begin{figure}
\includegraphics[width=0.49\columnwidth]{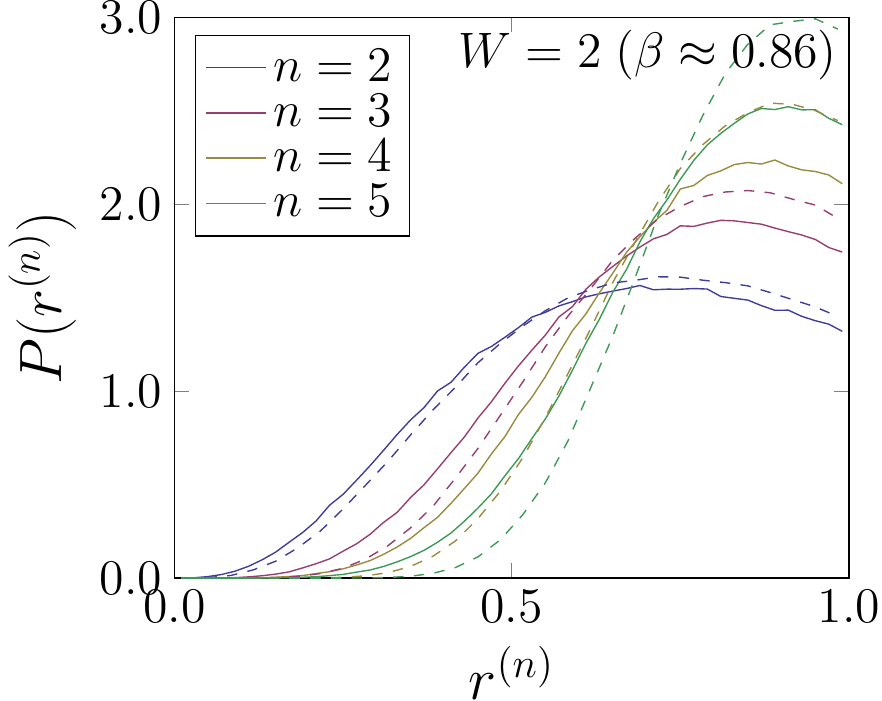} 
\includegraphics[width=0.49\columnwidth]{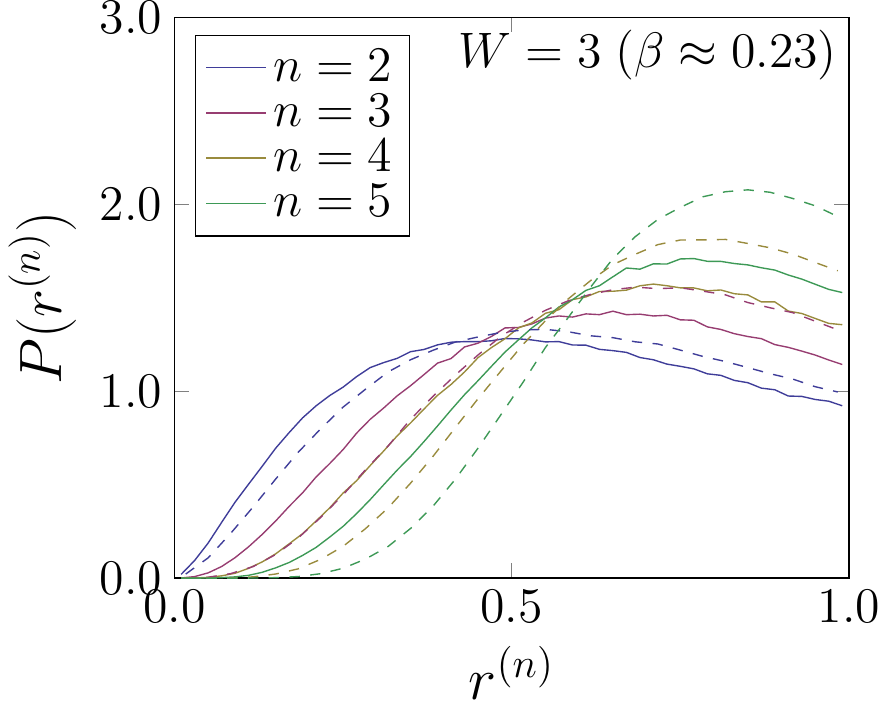} 

\smallskip

\includegraphics[width=0.49\columnwidth]{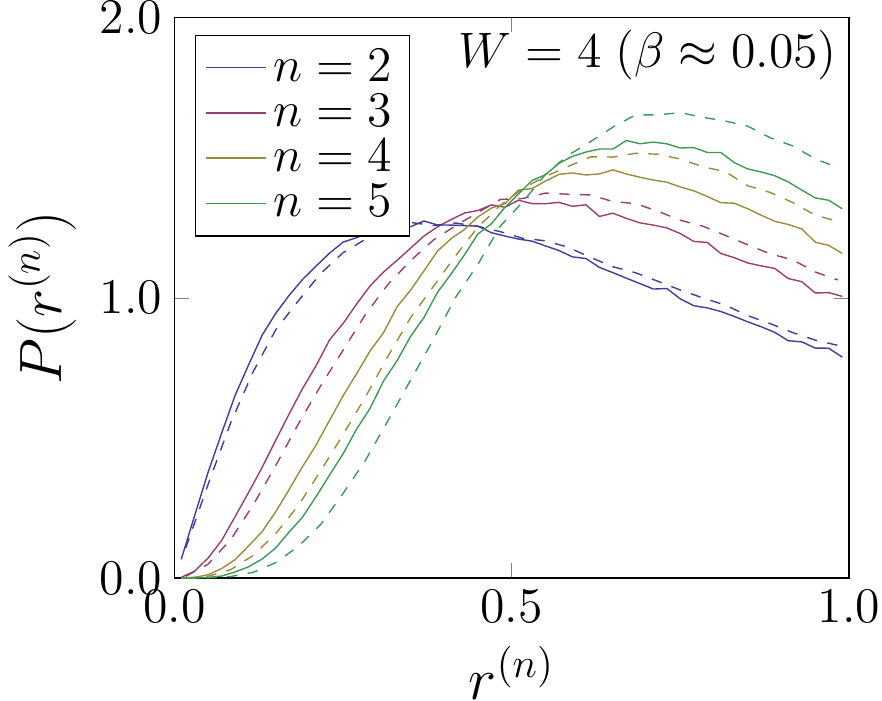} 
\includegraphics[width=0.49\columnwidth]{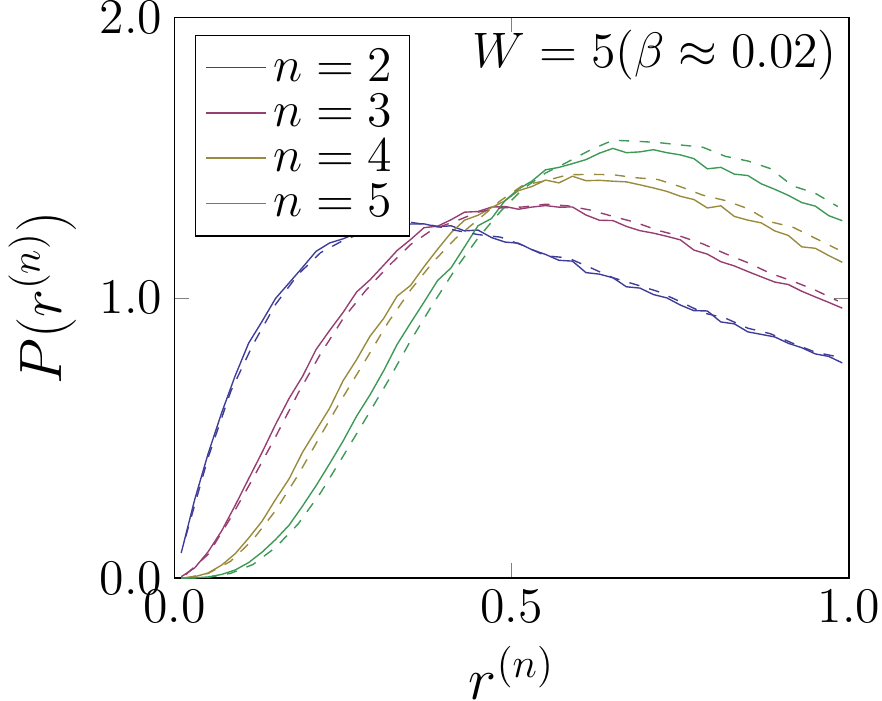} 

\caption{Numerically obtained distributions of $r^{(n)}$ with $n =2,3,4,5$ for the Hamiltonian at various $W$ (solid lines) and the corresponding distributions for the Gaussian $\beta$ ensemble (dashed lines, identical color scheme).}
\label{fig: corr}
\end{figure}

Fig. \ref{fig: corr} shows the distributions of $r^{(n)}$ for the Hamiltonian compared with the the corresponding distributions for the Gaussian $\beta$ ensemble, where the value of $\beta$ is estimated from $\langle r \rangle$. No unfolding is applied to the spectra of the Hamiltonian. We observe qualitative agreement up to $n=3$ (\emph{i.e.} up to $6$ level spacings) for all values of $W$. The algorithm used to unfold the spectra can be sub-optimal for the system under consideration. Attempts to compare the spectra of the Hamiltonian and the Gaussian $\beta$ ensemble on longer ranges by other measures such as the spectral rigidity \cite{Montambaux93}, density-density correlation function \cite{Joyner12}, and the spectral form factor \cite{Kos18} did not provide conclusive results, presumably due to this effect.

\textit{Discussion and conclusions}.--- 
We have proposed a generalization of Wigner-Dyson level statistics from the discrete taxonomy $\beta = 1,2,4$ to the continuous one $\beta \in (0, \infty)$. Using the matrix model for the Gaussian $\beta$ ensemble introduced in Ref. \cite{Dumitriu02}, we have shown how the Gaussian $\beta$ ensemble provides a smooth interpolation between Poissonian and Wigner-Dyson level statistics. We have studied the level statistics of a paradigmatic model in studies on MBL, and found near-perfect agreement with the corresponding statistics of the Gaussian $\beta$ ensemble over the full crossover range between the thermal (corresponding to $\beta \approx 1$) and many-body localized (corresponding to $\beta \approx 0$) phase, where $\beta$ is a single fitting parameter. We have shown that similar agreement holds for a related Hamiltonian with broken time-reversal symmetry.

We expect that this work paves a way for further investigations in various ways. Primarily, we believe it would be of significant interest to explore how universal the Gaussian $\beta$ ensemble describes the spectral statistics of quantum systems that show intermediate level statistics between Poissonian and Wigner-Dyson. In view of this, we note that there are several known physical and mathematical models supporting intermediate level statistics, studied mostly in the context of either single-particle models of quantum chaos \cite{Izrailev89, Prosen93, Prosen94, Bogomolny99} or the Anderson localization transition for non-interacting systems \cite{Moshe94, Kravtsov95, Kravtsov97, Sorathia12}. A crossover between Poissonian and Wigner-Dyson level statistics for $\beta = 2$ has also been found recently in a generalized SYK model \cite{GarciaGarcia18}.

Next, we expect that our results are of relevance in the field of MBL. In this field, level statistics are a key ingredient in both numerical \cite{Oganesyan07, Luitz15, Kjall18} and analytical \cite{Imbrie16} studies. The detailed quantitative characterization of the level statistics of the Hamiltonian provided in this work might be valuable in \emph{e.g.} the finite-size scaling analysis of the MBL transition \cite{Pal10, Luitz15} and studies on the intermediate phase separating the thermal from the MBL phase \cite{Khemani17} at finite system sizes. Finally, we hope that this work can contribute to the ongoing studies \cite{Muller04, Buijsman17, Kos18} on the fundamental correspondence between classical and quantum chaos.

\textit{Acknowledgements}.--- 
We thank Toma\ifmmode \check{z}\else \v{z}\fi{} Prosen and Maksym Serbyn for very useful discussions. V. G. acknowledges support from the Erwin Schr\"{o}dinger Institute in Vienna. This work is part of the Delta-ITP consortium, a program of the Netherlands Organization for Scientific Research (NWO) that is funded by the Dutch Ministry of Education, Culture and Science (OCW).

\bibliography{levels}

\end{document}